\documentclass[journal=jacsat,manuscript=article]{achemso}

\usepackage[version=3]{mhchem} 

\usepackage{pstricks}

\usepackage{epsfig}

\usepackage{psfrag}

\usepackage{endnotes}

\author{Xiulai Xu}
\email{xx757@cam.ac.uk}

\affiliation {Hitachi Cambridge Laboratory, Hitachi Europe Ltd., JJ Thomson Avenue, Cambridge CB3 0HE, United Kingdom}

\author{Hugh Baker}%
\affiliation{Microelectronics Research Centre, Cavendish
Laboratory, University of Cambridge, JJ Thomson Avenue, Cambridge CB3
0HE, United Kingdom}%

\author{David A. Williams}%
\affiliation{Hitachi Cambridge Laboratory, Hitachi Europe Ltd., JJ Thomson Avenue, Cambridge CB3 0HE, United Kingdom}%

\title{Highly sensitive, photon number resolving detectors mediated by phonons
using $\delta$-doped GaAs transistors}

\begin{document}







\begin{abstract}

 We report a photon number resolving detector using two-dimensional electron
gas (2DEG) based transistors. When the photon pulses impinge on the absorption
region, the generated phonons dissipate ballistically in the 2DEG toward the
trench isolated nanowire transistors near the surface. The phonon-electron
interaction induces a positive conductance in the transistors, resulting in a
current increase. With this principle, we obtain an internal quantum efficiency
for this type of detector of up to 85\%.

\end{abstract}





\maketitle

Highly sensitive photon detection is in demand for the implementation of
quantum information processing\cite{Takesue2007} and improved optical sensors
\cite{Konstantatos2006}. Recently, several photon number resolved (PNR)
detectors have been demonstrated using superconducting nanowires
\cite{Divochiy2008,Bell2007}, quantum dot gated field-effect transistors
\cite{Gansen2007}, superconducting transition-edge sensors
\cite{Miller2003,Rosenberg2005}, visible light photon counters \cite{Waks2004},
charge integration photon detectors \cite{Fujiwara2007}, silicon
photomultipliers \cite{Buzhan2006} and avalanche photodiodes
\cite{Kardynal2008}. Among them, there are two types of PNR detector. One
assembles multiple pixels to increase detection area
\cite{Buzhan2006,Divochiy2008} while the other only has a single detector
element \cite{Bell2007, Gansen2007,Kardynal2008}. In the first case, only the
photon number can be counted and it is not possible to resolve which pixel
detected the photons. In contrast, the single element detector gives the chance
to resolve photon energy and photon number when the photons impinge on the
absorber. In semiconductor-based optical detection, optical to electrical
conversion is normally realized by optically generated carriers in
semiconductors \cite{Clifford2009,Gansen2007,Kardynal2008,Fujiwara2007}. The
generated carriers either change the device current directly or act as a gate
through capacitive coupling. Because of the recombination of the carriers, the
sensitivity of the photon detection is intrinsically limited, which restricts
the use of such devices for high sensitivity applications. However, phonons are
also created with optical excitation \cite{Bron}, and it is possible to use
these optically generated phonons for photon detection.

When light with a photon energy larger than the band gap is absorbed by a
semiconductor, charge carriers are generated with a certain kinetic energy. The
excited electrons and holes then typically relax down to the conduction and
valence band edges via phonon emission. If the excess energy is larger than the
energy of one LO phonon, then LO phonons are emitted, rapidly decaying to
acoustic phonons within several picoseconds \cite{Baumberg1996}; otherwise,
only acoustic phonons are expected. When photon pulses impinge on an absorption
region, the generated phonons dissipate ballistically in the first instance.

In this letter, we report a photon number resolving detector, which comprises
an absorption region and two nanowire transistors, as shown in Figure 1 (a) and
(b). When photons impinge on the absorption area, phonons are generated and
then dissipated in a ballistic way towards the transistors aided by the phonon
focussing effect. The phonons reflected by the superlattice buffer layer, and
the back surface of the wafer, can be absorbed by electrons in the nanowires,
increasing their conductance. The quantum efficiency of this type of photon
detector can be up to 85\%.

When the phonons propagate towards the trench-isolated nanowire transistors
nearby (as shown in Figure 1 (a)), they can be absorbed by the transistors
across the trench because of the reflection from the superlattice layer and the
back surface of the wafer \cite{Mizuno2002,Kato1997}. Hot phonons have been
detected laterally using a 2DEG based nanowire transistor
\cite{Dzurak1992,Schinner2009}. Here we focus on the optically-generated phonon
assisted effect. In a two dimensional electron gas (2DEG) system, the transport
properties depend largely on electron-electron and electron-phonon
interactions. When the electron-electron interaction is much faster than the
electron-phonon processes, phonon absorption will increase the electron energy,
generally resulting in a conductivity change \cite{Blencowe1996,Blencowe1996b}.
$\delta$-doped GaAs has been demonstrated to be a good candidate of this type
of system \cite{Asche1992} because of the strong scattering by impurities.
Recently, a positive phonon-conductivity has been demonstrated  in a quasi-2D
$\delta$-doped GaAs based nanowire device, resulting in an increase (up to 40
$\mu$S/square) in the drain current of the transistor \cite{Poplavsky2000}.
With this principle, we expect optically-generated phonons will induce a
current increase, allowing photon detection.

A Si $\delta$-doped GaAs wafer with a carrier concentration around
$5\times10^{12}$ cm$^{-2}$, about 30 nm below the wafer surface, was used for
these optical detection devices. A superlattice buffer layer around 200 nm was
grown on the GaAs wafer to minimise defect propagation from the substrate into
the active region. A buffer layer of GaAs was then deposited with a thickness
of 1.5 $\mu$m before the $\delta$-doped layer growth. The device mesa was
patterned using standard electron-beam lithography and reactive ion etching
techniques. A layer of polymethylmethacrylate (PMMA) electron beam resist with
a thickness around 700 nm was used as an etch mask during the etching of the
isolation trenches. Ohmic contacts to the electrodes were formed by annealing a
deposited  Au:Ge:Ni alloy at $420\,^{\circ}{\rm C}$. The depth of the mesa
trench is about 1 $\mu$m. A cross section through the device is illustrated in
Figure 1 (c). Although the lithographic size of the wires is ~100 nm, surface
depletion means that the electrical width of the wire is much smaller, and can
be controlled by the lateral gate electrode.

The device was mounted onto a cold-finger cryostat, and measured at a
temperature of around 8.0 K. A DC bias from a low-noise voltage source in
series with low-noise filters was applied to each drain lead and to the gates
of both transistors. The gate voltages were optimized for sensitivity and
visibility. The outputs of the source electrodes were fed into two low-noise
current preamplifiers. A laser diode with emission wavelength at 808 nm was
driven with a dc bias and a superimposed voltage pulse from a function
generator. The frequency can be tuned from 0.01 Hz to 200 MHz. The power from
the laser was carefully calibrated and attenuated using neutral density
filters. Each neutral density filter was calibrated separately to obtain an
accurate light intensity. The variation of the stability of the laser diode
driver is around 0.5 \% and the error of the photon number calibration is about
1\%. The laser light after the attenuators was focused to the absorption area
using a long working distance objective ($\times$100). The beam spot is about
5 $\mu$m in diameter, as shown in Figure 1 (a). The reflection from the surface
of the absorption area is about 27.6 $\pm$ 2 \%, which was calibrated using a
beam splitter before the objective. This value corresponds well to the result
(28.3 \%) calculated by the Fresnel equation with a refractive index for GaAs
of 3.3. The average photon number referred to in this paper is the photon
number incident on the chip, excluding the reflection from the surface.

Figure 2 (a) shows the source-drain current of one of the transistors as a
function of gate voltage at different drain voltages. A typical transistor
characteristic can be observed and the drain current is saturated when the gate
voltage is around 0.5 V. This transistor can be operated as a single electron
transistor at 4.2 K or even in the optical cryostat at 8K, which means it is
very sensitive to changes in local charge distribution to well below the
single-electron level.

To demonstrate the phonon-induced conductance of the transistors, the
temperature-dependent conductance of one transistor is plotted in Figure 2 (b).
It can be seen that the conductance increases when the temperature increases
from 8 to 14 K. This indicates the positive phonon-induced conductance  (PIC)
in the nanowire transistors. Assuming the time of electron-phonon interaction
is much greater than the time of electron-electron collisions
\cite{Blencowe1996}, it is possible to observe positive PIC when the length of
a wire is longer than the elastic mean free path \cite{Kent1997}. This positive
PIC is due to the increase of electron temperature under the absorption of
phonon flux, which is given by \cite{Poplavsky2000}

\begin{equation}
{\frac{\delta\sigma}{G_{0}}\propto\frac{1}{T^{2}}\times[\frac{1}{e^{\hbar\omega_{q}/k_{B}T}-1}-\frac{1}{e^{\hbar\omega_{q}/k_{B}T_{0}}-1}]},
\end{equation}

where $G_{0}=e^{2}/(2\pi^{2}\hbar)$, $k_{B}$ is the Boltzmann constant and
$T_{0}$ is the temperature before heating. The solid line (red) shows the
fitted results in Figure 2 (b). In this case, the electron temperature is same
as the crystal temperature. It should be noted that for the weak light detection,
especially for the photon number resolved detection in this work, the photon
induced crystal temperature change can be neglected, being much less than 1 mK.

Further evidence for an optically generated phonon-induced positive current can
be found by biasing the two transistors relative to one-another. The two
transistors were floated and a bias voltage was applied between them. With this
lateral field, the electron-hole pairs generated by optical pumping will be
separated; electrons moving to one transistor and holes to the other. The top (bottom) panel in Fig. 2 (c) shows the optical response of the low
(high) potential transistor. At the transistor with low (high) potential, holes
(electrons) move towards it and act as a positive (negative) gate voltage,
resulting in a current increase (decrease) at low optical power. For the
transistor at high potential, the current decreases only at low optical power
(1\% of the laser power, the bottom trace). When the optical power is increased
to 10\%, the current decreases quickly due to the negative charge gating, then
increases slowly, which we ascribe to the heating effect. With a power around
30\%, no negative gate effect can be observed.

To perform single-shot measurements, the laser diode was modulated at 1 Hz, and
the drain currents were measured by a voltage meter via low-noise current
pre-amplifiers. Figure 3 (a) shows the single-shot measurements of photon
induced current of one of the transistors with an average photon number
($\overline{\mu}$) in each pulse  of $8.9\times 10^6$ and 712, respectively. The
current builds up within the optical pulse up to about 2 $\mu$A and then
decreases slowly. When $\overline{\mu}$ decreases, the amplitude of the current
peaks reduces. With $\overline{\mu}$ at $8.9\times 10^6$, the current continues
to increase about 50 ms after the optical pulse is switched off, then
decreases. This slower response, compared with the results in ref [22], is due
to the low-pass filters in this measurement. It should be noted that the
current did not drop to the equilibrium level at high optical power before the
next pulse arrived, therefore the current increase does not linearly depend on
the power. The responses of the two transistors are similar. The current change
can be measured when the laser intensity was attenuated to a few photons per
pulse. The real-time signal with ($\overline{\mu}$) from 1.6 to 8.2 per pulse
is shown in Figure 3 (b). We concentrate on low photon number detection here
because the conductance does not always respond linearly with phonon number
\cite{Poplavsky2000}.

In order to confirm photon number resolution, histograms of the current change
of the two transistors were constructed, as shown in Figure 4. Figure 4 (a)-(d)
and (e)-(h) show the histograms of Transistor 1 and Transistor 2 respectively, with average
photon number $\overline{\mu}$ of 0, 0.63, 1.15, and 1.69. It can
be seen that each number of photons can be well resolved. The solid blue line
is the sum of a series of Gaussian functions with the same peak separation.
The average current increase for one photon step is approximately $0.50\pm0.05$
nA. The photoconductive responsivity of these devices is around $2.6\times10^9$
A/W. The overall envelope current histogram in each $\overline{\mu}$ shows a
Poisson distribution, as would be expected from the photon number distribution
in laser pulses. Figure 4 (i)-(l) shows the calculated Poisson distribution
with average photon number of 0, 0.50, 0.99, 1.35, which corresponds to the
$\overline{\mu}$ when considering the quantum efficiency of 80 \%. It can be
seen that the observed results are in accordance with what was expected from
theory. The intensities of 0 and 1 photon peaks in Figure 4 (c) and (g)
slightly deviate from the theory shown in Fig. 4 (k), which may be due to the
laser spot being slightly off-centre. Because of the overlap between two
adjacent Gaussian peaks, the accuracy of determining the photon number is
around 85$\pm$5 \%.

By counting the photon numbers measured within the each number state bin, and
considering the Poisson distribution of photons from a laser, the internal
quantum efficiency can be calculated. Figure 5 shows the internal quantum
efficiency as a function of each photon number state. It can be seen that the
quantum efficiency varies from 70 \% to 85 \% with an error bar about 5 \%, and
does not change very much up to 4 photons. This internal quantum efficiency is
comparable to that of charge integration photon detectors \cite{Fujiwara2007}
and slightly lower than that of the superconducting transition-edge sensor
\cite{Rosenberg2005}. It should be noticed that the quantum efficiency here
represents the proportion of detected photons for a given number of absorbed
photons. This is different from the efficiency of optically generated phonons
detected by the transistors: The photon detector counts once when there are
phonons absorbed by the transistor, which can be a small percentage of the
phonons generated by each photon.

The dark count rate is about 0.08 per second with a discrimination level of
0.25 nA, which is due to the noise from the nanowire devices without
passivation. For PNR detectors, the noise equivalent power (NEP) is usually
defined as $NEP=(h\nu/\eta)\sqrt{2R}$ \cite{Miller2003}, where $h\nu$ is the
photon energy, $\eta$ is the detection efficiency and $R$ is the dark count
rate. It gives a NEP for our devices of about $1.23\times10^{-19}
W/\sqrt{Hz}$. The current responsivity is about 2.6 times of that of quantum
dot gated field-effect transistors for the detection of one photon
\cite{Gansen2007}. We noticed that the NEP can be as low as $10^{-21}
W/\sqrt{Hz}$, when the light is shining directly on one transistor. However, a
single transistor device is not practical for photon detection because of the
small absorption area. This high sensitivity is due to the following reasons:
At a low temperature (less than 10 K), the generated acoustic phonons dissipate
ballistically at the speed of sound \cite{Kent1996}.  Because of the anisotropy
of the GaAs crystal structure, the phonon phase and group velocities are not
colinear, resulting in a phonon focusing effect \cite{Taylor1971,Bron,Wolfe}.
Karl \textit{et al.} \cite{Karl1988} observed the phonon focusing pattern on
the 2DEG by measuring the phonon-drag voltage pattern. When light impinges on
the 2DEG, the generated phonons propagate along the \{100\} planes and \{110\}
planes, which correspond to the fast transverse phonons and slow transverse
photons, respectively. In this work, the detection devices were patterned along
the [100] direction, as shown in Figure 1 (a).

Another reason for this high responsivity might be due to resonant excitation.
For Si $\delta$-doped GaAs, the main luminescence peak is around 830 nm at 8 K,
which is ascribed to a donor to Si acceptor transition. The energy difference
between the laser diode (808 nm) and the main luminescence peak is around 37
meV, which is similar to the optical phonon energy (around 36 meV) in GaAs. The
phonon emission can be enhanced at this resonant condition because of the
resonant enhancement of inelastic light scattering by the 2DEG
\cite{Danan1989,Nakayama2007}.

In conclusion, we have demonstrated an ultra-high sensitivity photon detector
using $\delta$-doped GaAs based devices, mediated by phonons. The photon number
can be resolved up to 4 photons with an accuracy of 85$\pm$5 \%. The
responsivity is around $2.6\times10^9$ A/W, where the NEP is about
$1.23\times10^{-19}  W/\sqrt{Hz}$. The internal quantum efficiency is measured
to be around 80\%. Overall, we believe this phonon mediated optical sensor
offers a new approach for highly sensitive photon detectors using
semiconductor-based devices.

\begin{acknowledgement}

We acknowledge Dr Andrew Armour and Dr Hongwei Li for very helpful discussions
and comments.

\end{acknowledgement}

\newpage
\bibliography{photondetection}

\newpage

 \begin{figure*}

\centering

\epsfig{file=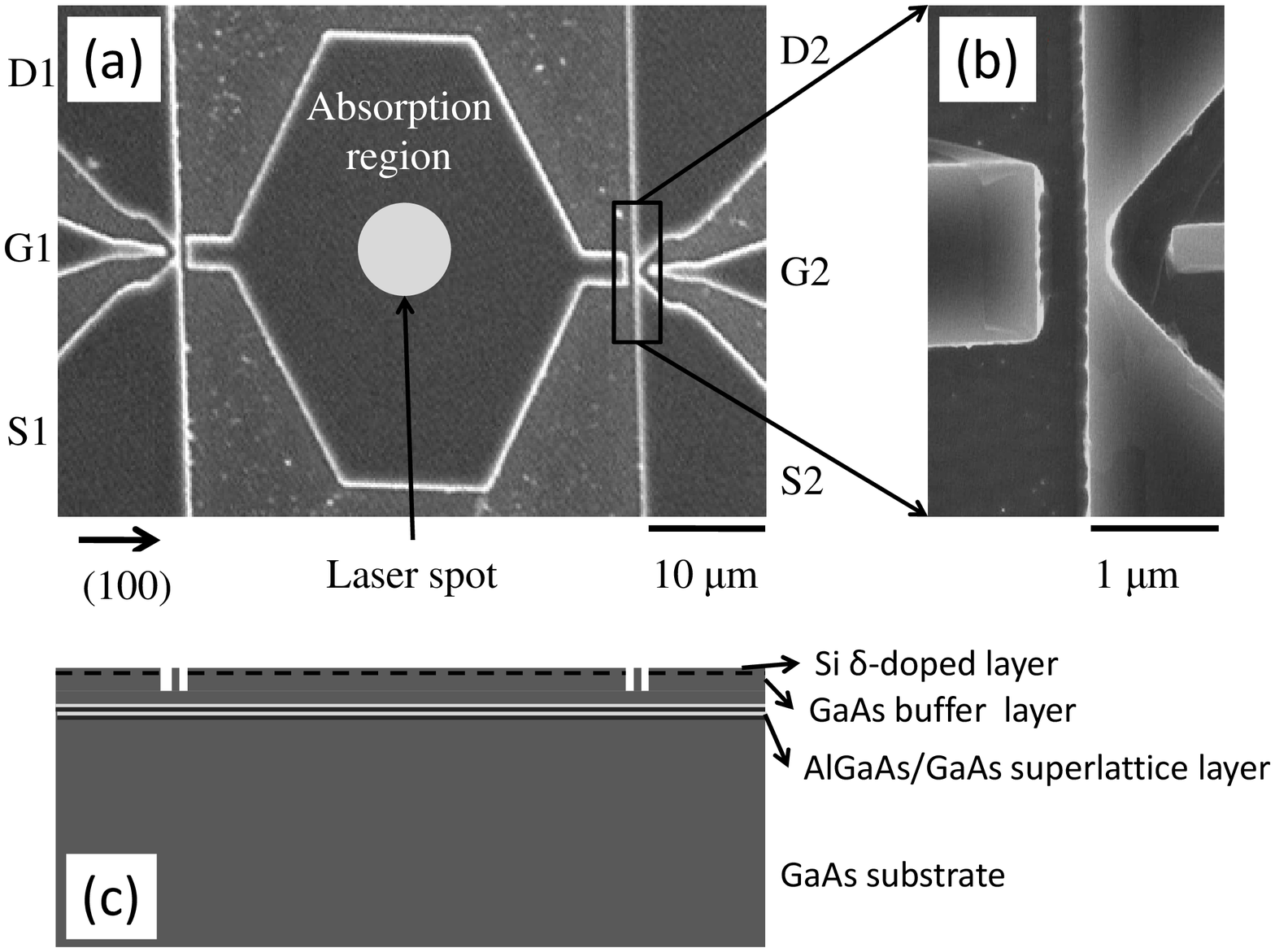,width=12cm,keepaspectratio}

\caption{\label{fig:figure1} (a) A SEM image of a photon detection device. Two
trench isolated nanowire transistors were separated with a distance of around
30 $\mu$m, between which the absorption area is located. The laser was focused
on the central region of the absorption area (bright spot). (b) An enlarged SEM
image of one of the transistors. The width of the nanowire is about 200-300 nm,
with a length of 800 nm. The separations between gate/absorption region and the
nanowire channel are around 500 nm. (c) A cross-section sketch along the centre
of the device in (a). }

\vspace{5cm}

\end{figure*}

 \begin{figure*}

\centering

\epsfig{file=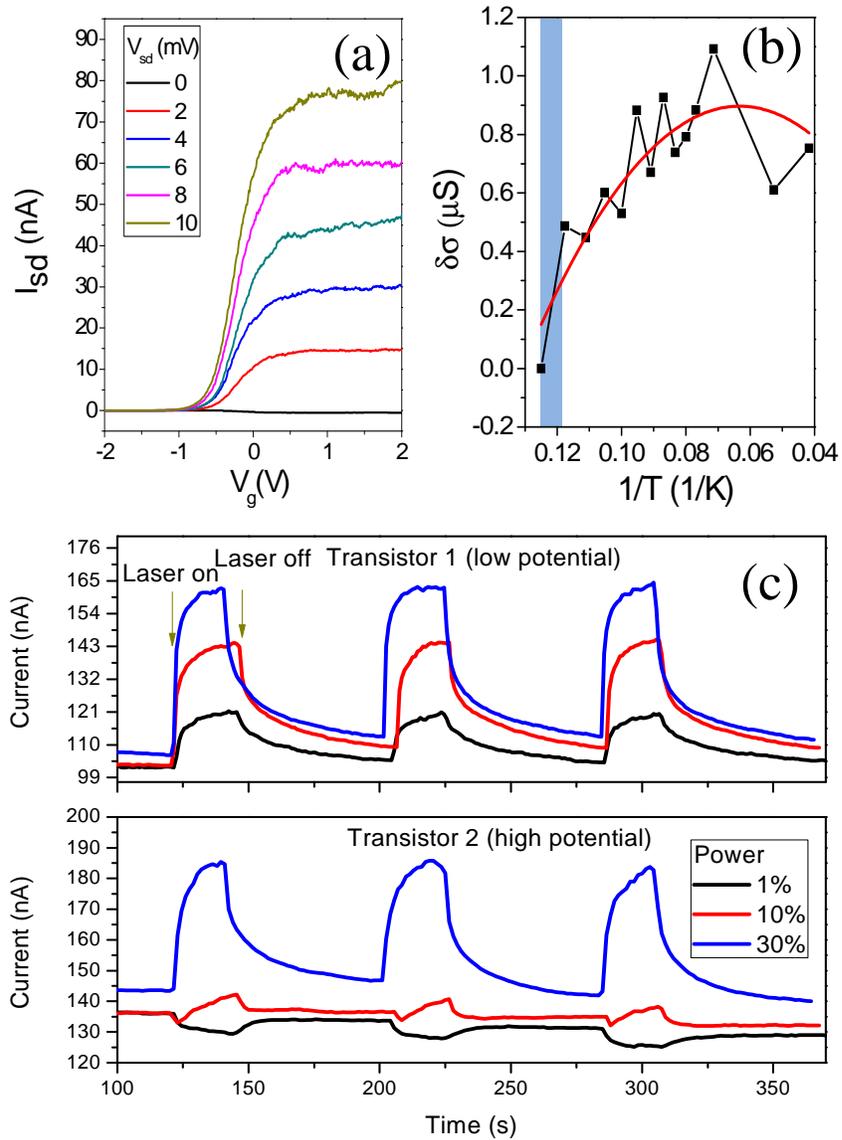,width=12cm,keepaspectratio}

\caption{\label{fig:figure2} (a) A typical transistor drain current as a
function of gate voltage for different source-drain voltages. (b) Conductance
of one nanowire transistor as a function of substrate temperature. The solid
line (red) shows the fitted result with the phonon induced conductivity theory.
(c) Optical-induced current change with different optical intensities with a
biased voltage between two transistors. }

\vspace{5cm}

\end{figure*}

\begin{figure*}

\centering

\epsfig{file=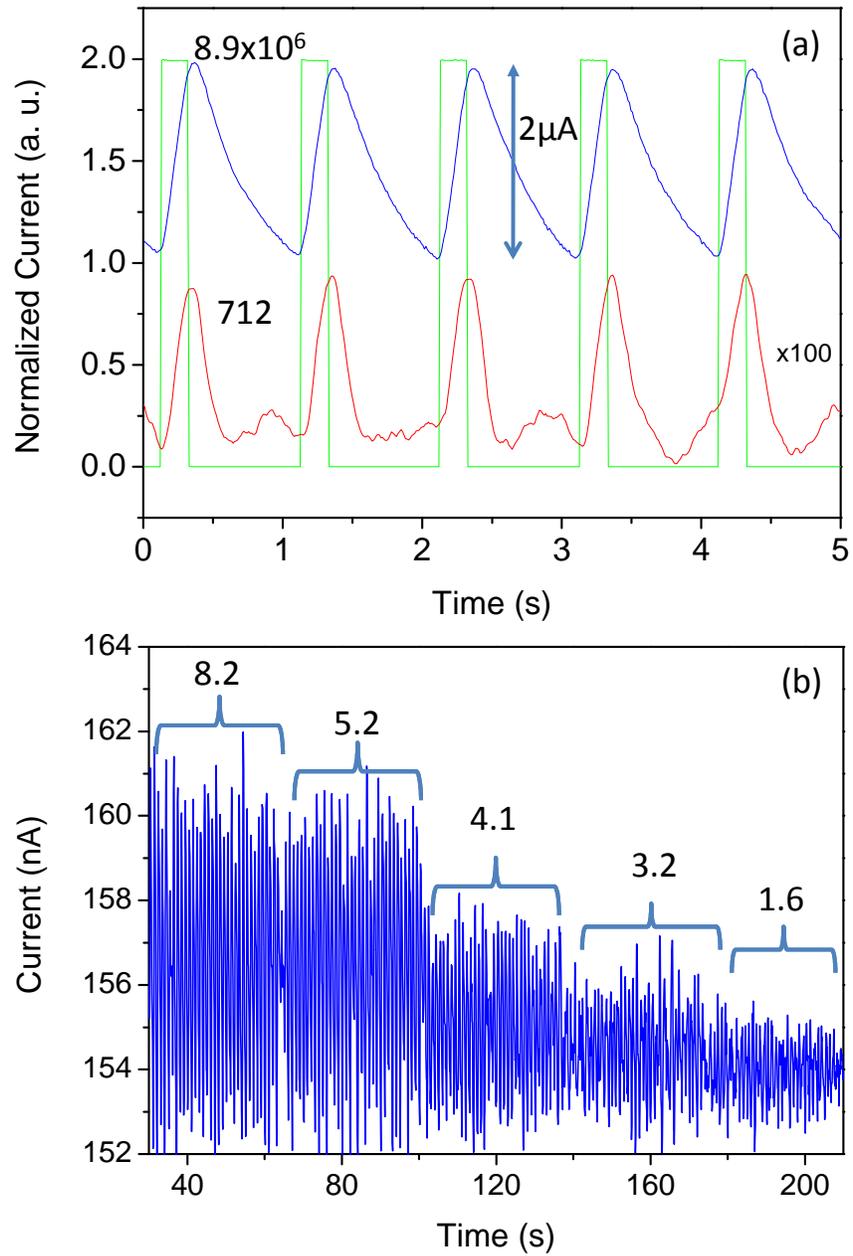,width=12cm,keepaspectratio}

\caption{\label{fig:figure3} (a) Normalized drain current changes of one of the
two transistors with average photon numbers of $8.9\times 10^6$ (top trace) and
712 (bottom trace) respectively at 1 Hz. The green rectangular trace shows the
SYNC signal from the laser diode driver. The current traces are normalized to
the maximum current intensities and shifted for clarity. (b) The real-time signal of current increase with
($\overline{\mu}$) varying from 8.2 to 1.6. }

\vspace{5cm}

\end{figure*}

\begin{figure*}

\centering

\epsfig{file=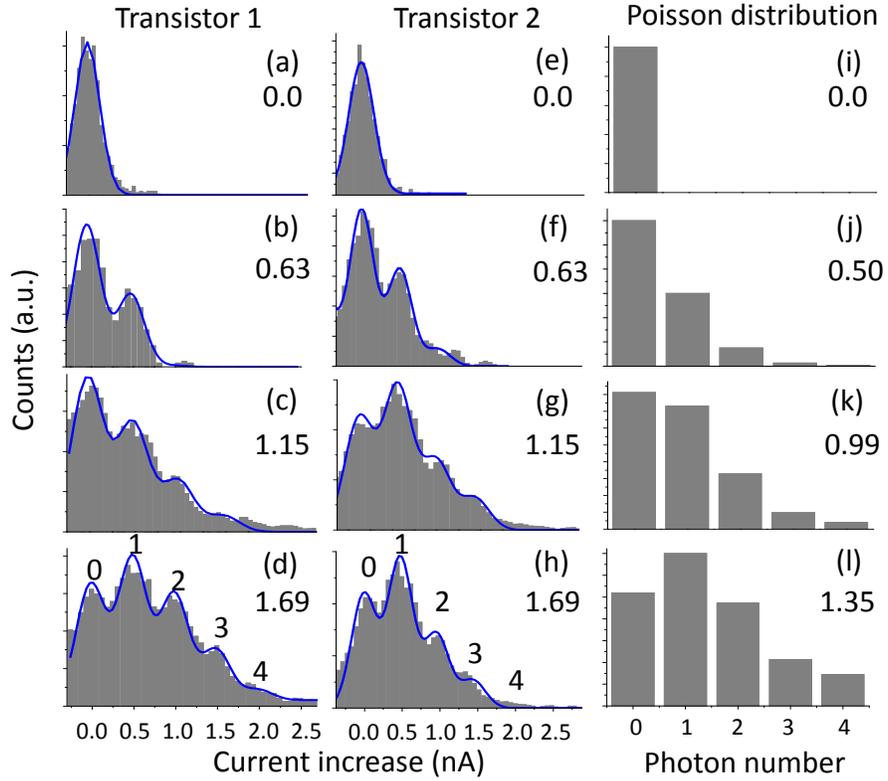,width=12cm,keepaspectratio}

\caption{\label{fig:figure4} Histograms of detected photon number corresponding
to the binned current changes of the two transistors with $\overline{\mu}$ at
0, 0.63, 1.15 and 1.65 (a)-(h). The solid blue line is the envelope of the
fitted Gaussian peaks in each panel. The photon number states are marked in (d)
and (h). (i)-(l) are calculated Poisson distribution of $\overline{\mu}$ at 0,
0.5, 0.99 and 1.35, which are corrected average number from (a)-(h) with an
internal quantum efficiency at 80 \%.}

\end{figure*}

\begin{figure*}

\centering

\epsfig{file=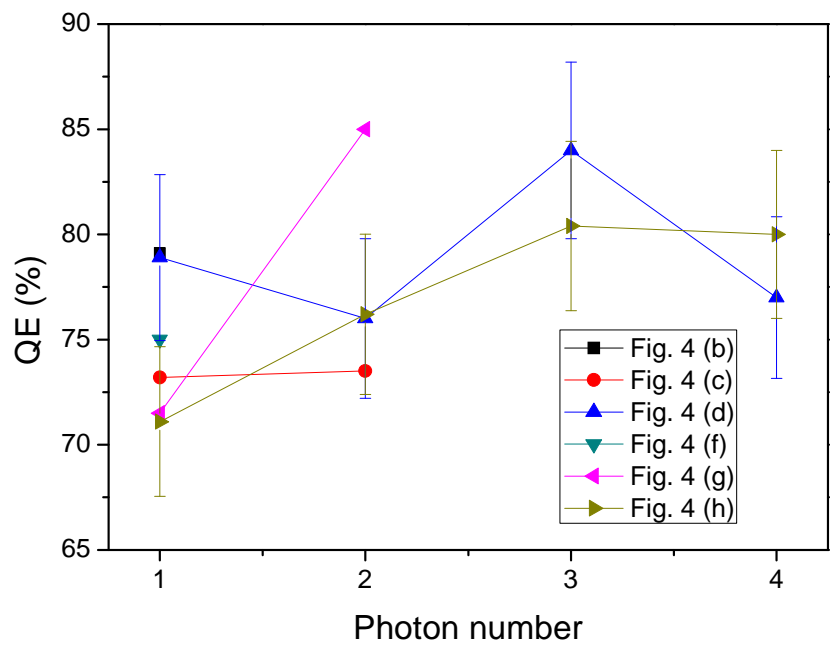,width=12cm,keepaspectratio}

\caption{\label{fig:figure5} Internal quantum efficiencies, corresponding to
the measurements in Figure 4, as a function of photon number state. }

\end{figure*}

\end{document}